\documentclass[12pt]{article}
\usepackage{amsmath,amssymb}
\newcommand{\beq}{\begin{equation}}
\newcommand{\eeq}{\end{equation}}
\newcommand{\ba}{\begin{array}}
\newcommand{\ea}{\end{array}}
\newcommand{\bea}{\begin{eqnarray}}
\newcommand{\eea}{\end{eqnarray}}
\newcommand{\bean}{\begin{eqnarray*}}
\newcommand{\eean}{\end{eqnarray*}}

\newtheorem{theorem}{Theorem}[section]
\newtheorem{prop}[theorem]{Proposition}

\newtheorem{defi}[theorem]{Definition}
\newtheorem{remark}[theorem]{Remark}
\newenvironment{rem}{\begin{remark} \rm}{\end{remark}}
\newtheorem{proof}{Proof.}
\newenvironment{pf}{\begin{proof} \rm}{\hfill$\square$ \end{proof}}

\newcommand{\CZ}{{\cal Z}}
\newcommand{\CS}{{\cal S}}

\newcommand{\CM}{{\cal M}}

\newcommand{\CL}{{\cal L}}

\makeatletter
\@addtoreset{equation}{section}

\makeatother

\newcommand{\CC}{{\mathbb C}}

\newcommand{\ZZ}{{\mathbb Z}}

\def\la{\lambda}

\newcommand{\lmp}[3]{Lett. Math. Phys. {\bf #1} (#2), #3}

\newcommand{\TT}{{\mathsf T}}
\newcommand{\FF}{{\mathsf F}}
\newcommand{\rref}[1]{(\ref{#1})} 

\newcommand{\del}{{\partial}}

\newcommand{\Haa}[1]{H^{(#1)}}
\newcommand{\vefi}{{vector field}}

\def\a#1{a^{(#1)}}

\def\ger{hierarch}
\def\var{manifold}

\def\bih{bihamiltonian}
\def\varb{\bih\ \var}
\def\ham{Hamiltonian}

\def\ger{hierarch}

\def\varb{\bih\ \var}

\newcommand{\wid}[1]{\widehat{#1}}
\begin{document}
\begin{flushright}
Ref. SISSA 139/1999/FM
\end{flushright}
\vspace{0.8truecm}
\begin{center}
{\huge Bihamiltonian geometry and  separation of 
variables  for Toda  lattices}
\end{center}
\vspace{0.8truecm}
\makeatletter
\begin{center}
{\large
Gregorio Falqui${}^1$,
Franco Magri${}^2$, and
Marco Pedroni${}^3$}\\ \bigskip
${}^1$ SISSA, Via Beirut 2/4, I-34014 Trieste, Italy\\
E--mail: falqui@sissa.it\\
${}^2$ Dipartimento di Matematica e Applicazioni\\ 
Universit\`a  di  Milano--Bicocca \\
Via degli Arcimboldi 8, I-20126 Milano, Italy\\
E--mail: magri@vmimat.mat.unimi.it\\
${}^3$ Dipartimento di Matematica, Universit\`a di Genova\\
Via Dodecaneso 35, I-16146 Genova, Italy\\
E--mail: pedroni@dima.unige.it
\end{center}
\makeatother
\vspace{0.1truecm}
\noindent {\bf Abstract}. 
We discuss the \bih\ geometry of the Toda lattice
(periodic and open). Using some recent results on the separation
of variables for \varb, we show that these systems can be 
explicitly integrated via the classical Hamilton--Jacobi method in 
the so--called Darboux--Nijenhuis coordinates. 
\vspace{.1truecm}

\section{Introduction}\label{sect:intro}
In this paper we reexamine the classical $A_{n}$--Toda systems with the  aim
of showing that these lattices fall into a notable class of \bih\ 
integrable systems: those for which a distinguished set of coordinates (the
so--called {\em Darboux--Nijenhuis (DN)} coordinates) allows the solution of
the Hamilton--Jacobi equations associated with the Hamiltonian flows by means
of an (additive) separation of variables (SoV)~\cite{Sk,Ha1}. In particular,
we will show that such coordinates arise from the geometry of 
the Poisson pencil after a Hamiltonian
reduction process on  suitable symplectic leaves.

DN coordinates (see, e.g., \cite{Ma90}) 
can be naturally defined on a Poisson--Nijenhuis (PN) manifold \cite{KM}, 
that is, on a $2n$--dimensional  manifold $\CM$ endowed with a 
symplectic two--form $\omega$ and a $(1,1)$ torsion free tensor 
$N$ satisfying certain compatibility conditions.
In~\cite{fmp2} we 
present and discuss an intrinsic condition  
to characterize those  
\ham\ systems on $\CM$ for which DN 
coordinates separate the corresponding 
Hamilton--Jacobi equations.
Moreover, some of the connections between Hamiltonian \ger ies
which satisfy a certain recursion property 
with respect to the tensor $N$,  
and Gel'fand--Zakharevich (GZ) systems~\cite{GZ93} 
are investigated there. 
This paper is devoted to frame the Toda lattices into such a
scheme. 

The plan is as follows: in Section~\ref{sect:one} we sketch the main
points of the abovementioned SoV theory for \varb, referring to~\cite{fmp2}
for complete proofs and a more detailed discussion.  
Section~\ref{sect:two} contains a formulation of the Toda lattice 
within the GZ scheme, 
that is, 
taking as starting point its Poisson 
pencil and the problem of finding the Casimir functions.
Section~\ref{sect:three} concerns the application of the \bih\ SoV theory to
this family of integrable systems. Finally, in Section~\ref{sect:four} we
treat the three-particle case to give a feeling of 
how the method works. 
  
\section{Separation of variables on PN manifolds}\label{sect:one}
Let $(\CM,\omega)$ be a $2n$--dimensional 
symplectic manifold endowed with a Nijenhuis tensor field $N$ compatible with
$\omega$ (in the sense of the theory of \varb s). These manifolds are called
Poisson--Nijenhuis manifolds~\cite{KM}. Examples of such manifolds are
provided by \varb s endowed with a pair of Poisson bivectors $(P_0,P_1)$ 
one of which, say $P_0$, is invertible, In this case, 
$\omega=P_0^{-1}$ and $N=P_1 P_0^{-1}$.
\begin{defi}
\label{defi:dncord}
By {\em Darboux--Nijenhuis} coordinates on $\CM$ we mean any system of local
coordinates $(\lambda_j,\mu_j)_{j=1,\ldots,n}$ which enjoy the following two 
properties:\\
i) $\omega$ takes the canonical form 
\[
\omega=\sum_{i=1}^n d\la_i\wedge d \mu_i\ ;
\]
ii) the adjoint Nijenhuis operator $N^*$ takes the diagonal form 
\[
N^* d\la_j=\la_j d\la_j\ ,\qquad N^* d\mu_j=\la_j d \mu_j\ .
\]
\end{defi}
It has been shown~\cite{Ma90} that DN coordinates exist on any PN manifold
where   $N$ has $n$ functionally independent eigenvalues. In this
case the coordinates   $\lambda_j$ can be computed algebraically
as  the roots of the minimal polynomial of $N$, 
\begin{equation}\label{eq:lacoord}
C(\lambda)=\mbox{Det} \big(N-\lambda\mathbf 1\big)^{\frac12}\ .
\end{equation}
On the contrary, the complementary coordinates 
must be computed  (in general)  by  a method involving  quadratures. 

In~\cite{fmp2} we characterize a class of Hamiltonians on $\CM$ whose
associated Hamilton--Jacobi equations can be solved by separation 
of variables in DN coordinates. 
Let $(H_1,\ldots,H_n)$ be a set of functionally independent
(Hamiltonian) functions that are
in involution with respect to the canonical Poisson bracket defined by
\[
\{f,g\}=\omega(X_f,X_g)\ .
\]
We {\em assume} that the Lagrangian foliation defined by the functions $H_i$
is invariant with respect to 
$N$. This is tantamount to saying that, at a generic 
point $m\in \CM$, the differentials  $dH_k$ span 
an $n$--dimensional vector subspace of $T^*_m \CM$ which is closed under 
the action of $N^*$. Hence there exists  
an $n\times n$ matrix $\FF$, whose entries are functions on $\CM$, such that
\begin{equation}\label{eq:hfeq}
N^* d H_i= \sum_{j=1}^n \FF_i^j d H_j\ .
\end{equation}
\begin{defi}\label{def:separa} The Hamiltonians $(H_1,\ldots,H_n)$
are separable 
in the DN coordinates if there exists an $n\times n$ invertible matrix $\TT$
and an $n$--component vector $V$ such that
\begin{equation}\label{eq:stackeq}
\TT H=V\ ,
\end{equation}
where $H=(H_1,\ldots,H_n)^T$, and the matrix $\TT$ and the vector $V$ possess
the St\"ackel properties:
\begin{enumerate}
\item
the entries of the $j^{th}$ row of $\TT$ depend only on the conjugated
coordinates $(\la_j,\mu_j)$.
\item  the $j^{th}$ component of the vector $V$  depends only 
on $(\la_j,\mu_j)$ as well.
\end{enumerate}
\end{defi}
A remarkable ``separability test'' is given by the following
\begin{theorem}\label{teo:feq}
The Hamiltonians  $(H_1,\ldots,H_n)$ are separable
if and only if the matrix $\FF$ verifies 
the equation 
\begin{equation}\label{eq:feq}
N^*d\FF=\FF d\FF\ .
\end{equation}
\end{theorem}
Two remarks are in order to explain this theorem. 
{}First of all, equation~\rref{eq:feq} must be
read as follows. In the left hand--side, $d\FF$ is the matrix whose 
entries are the differentials of the entries of $\FF$, 
and $N^*$ acts separately on each entry.
Secondly, one should notice that~\rref{eq:feq} is a {\em coordinate free} 
test of separability, that can be checked without 
computing the DN coordinates.
Once the test is passed one can construct the St\"ackel matrix $\TT$, still in
general coordinates, by a simple algebraic procedure. One
has to consider the eigenvectors of the matrix $\FF$ and form with them a
(suitably normalized) matrix $\TT$ that diagonalizes $\FF$:
\[
\FF=\TT^{-1} \Lambda \TT\ ,\qquad 
{\Lambda}=\mbox{ diag }(\lambda_1,\ldots, \lambda_n)\ .
\]
By condition~\rref{eq:feq}, this matrix is a St\"ackel matrix; 
by condition~\rref{eq:hfeq}, the vector $V=\TT H$ verifies the St\"ackel
property.
Then, once constructed the DN coordinates, the Hamilton--Jacobi 
equations associated with 
$(H_1\ldots,H_n)$ can be easily solved by separation of variables. Notice that
the DN coordinates separate at once the HJ equations associated with {\em
any} of the Hamiltonians $H_i$.

To complete the construction of the DN coordinates, 
that is, to construct algebraically the coordinates $\mu_j$
conjugated to the eigenvalues $\la_j$ of $N$, 
the following procedure is often
useful. We consider the Hamiltonian vector 
field $Y$ associated (by the symplectic form $\omega$)
with the function $\frac12 \mbox{Tr}(N)$,
and the space of  functions $F(x;\la)$, depending smoothly   on $x\in \CM$  
and  holomorphically  on the parameter $\lambda$. 
We denote with $F(x;\lambda_j)$ the evaluation of
$F(x;\lambda)$ at $\lambda=\lambda_j$.
If ${N}^* dF(x;\la_j)=\la_j dF(x;\la_j)$ 
for all $j=1,\dots,n$, we say that $F(x;\lambda)$ is an 
{\em exact eigenvector} of $N^*$. 
\begin{theorem}\label{prop:y}
If $F(x;\la)$ is an exact eigenvector of $N^*$, satisfying the
``normalization property''  $Y(F(x;\la))=1$, then
the {\em evaluation} of 
$F(x;\lambda)$ at the points $\lambda=\lambda_j$, i.e.,
\[
\mu_j=F(x ; \la_j)\ ,
\]
provides a set of $n$ remaining DN coordinates.
\end{theorem}
In the application of Section~\ref{sect:three}, we will use the
property 
that if $F(x;\lambda)$ is an exact eigenvector, then $Y(F(x;\lambda))$
is an exact eigenvector as well. Since in the separable case a suitable
combination of the Hamiltonians
is exact, one can act with $Y$ on such a combination
and generate a space of exact eigenvectors where the equation
$Y(F(x;\la))=1$ may be solved algebraically. 

\subsection{DN separable Hamiltonians from GZ systems}\label{sect:2_1}
Let  $\CM$ a $(2n+k)$-dimensional manifold endowed with a pencil $P_\la=
P_1-\la P_0$ of Poisson tensors. We suppose that it admits $k$ polynomial
Casimir functions
\[
\Haa{a}=\sum_{j=0}^{n_a}\Haa{a}_j\lambda^{n_a-j}\ ,\quad
a=1,\ldots,k\ ,
\]
with  $n=n_1+\cdots+n_k$.
If the functions $\Haa{i}_j$ are functionally independent, then
$\CM$ is  called a {\em complete} GZ manifold, and the pencil
$P_\la$ is said to be a pure Kronecker pencil of 
type $\{2n_1+1,\ldots,2n_k+1\}$. Since the functions 
${H}^{(a)}_{0}$ form a maximal set of independent Casimirs of $P_0$,
the 
generic symplectic leaf $\CS$ of $P_0$ is the $2n$--dimensional 
submanifold given by 
${H}^{(a)}_{0}=C_a$, for $a=1,\ldots, k$. The restrictions 
$\wid{H}^{(a)}_{j_a}$  to $\CS$, for $j_a=1,
\ldots, n_a$, and $a=1,\ldots, k$, of the $n$ remaining Hamiltonians
define a completely integrable system in the Liouville sense. 

In order to solve by SoV this system, we
suppose that there exist $k$ vector fields $Z_a$ 
(to be called {\em transversal\/} vector fields) spanning a
$k$--dimensional integrable distribution $\CZ$ and satisfying:
\\a) The normalized transversality condition: 
$\mbox{Lie}_{Z_a}(\Haa{b}_0)=\delta^b_a$ for all $a,b=1,\dots,k$;
\\ b)  The deformation condition for the Lie derivatives: 
$\mbox{Lie}_{Z_a}  (P_\la)=\sum_{b=1}^k Z_b\wedge Y^b_a$ for 
some vector fields $Y_b^a$;
\\ c) The ``flatness'' condition:
$\mbox{Lie}_{Z_a}(\mbox{Lie}_{Z_b}(\Haa{c}(\la)))=0,\> \forall a,b,c$.\\
Conditions a) and b) imply that the distribution ${\cal Z}$ is
transversal to the symplectic leaves of $P_0$, and that the
functions vanishing along $\CZ$ are a {\em Poisson subalgebra\/} with 
respect to the Poisson pencil $P_\la$. 
Then, as a consequence of the Marsden--Ratiu
theorem~\cite{MR}, we have that:
\begin{prop}\label{prop:g}
The Poisson pencil on $\CM$ can be projected on 
the generic symplectic leaf $\CS$ of $P_0$, so that $\CS$ becomes 
a PN manifold. The functions $\wid{H}^{(a)}_{j_a}$, 
for $j_a=1, \ldots, n_a$, and $a=1,\ldots, k$, 
satisfy the condition~\rref{eq:hfeq}, that is, there exists
an $n\times n$ matrix $\FF$ such that
$N^* d \wid{H}=\FF \wid{H}$, where $\wid{H}$ is a column vector collecting the
above functions.
\end{prop}
Under the ``flatness'' condition c), one can show that 
equation \rref{eq:feq} is satisfied, 
so that the reduced Hamiltonian system is separable in the DN coordinates.
These coordinates may be computed from the geometry of $P_\la$,
without actually performing the reduction process.
In this case, in fact:
\begin{enumerate}
\item The minimal polynomial 
of the Nijenhuis tensor $N$ induced, according to the previous proposition, on 
the leaf $\CS$ is the determinant of the matrix
$G(\la)=\left[\mbox{Lie}_{Z_a}(\Haa{b}(\lambda))\right]_{a,b=1,\dots,k}$, 
that is, $\det G(\la)=0$ iff $\la=\la_j$;
\item The vector field $Y$ of 
Theorem~\ref{prop:y} is given by 
$Y=\sum_{a=1}^k Y^a_a$;
\item If $(1,\rho_2(\la),\dots,\rho_k(\la))$ satisfies 
\[
(1,\rho_2(\la),\dots,\rho_k(\la))G(\la)=0\qquad \mbox{for $\la=\la_j$,}
\]
then
$\Haa{1}(\la)+\rho_2(\la)\Haa{2}(\la)+\dots+\rho_k(\la)\Haa{k}(\la)$
is an exact eigenvector of $N^*$. Hence it can be used to find a
normalized exact eigenvector, and therefore the $\mu_j$ coordinates.
\end{enumerate}

\begin{rem}
The SoV theory for PN manifolds outlined above
provides intrinsic and algorithmic recipes to check whether a given
Liouville integrable system defined on a PN manifold can be separated in
the DN coordinates. On the other hand, the conditions under which  
one obtains separable Hamiltonians from a GZ manifold are by no means
algorithmic. In particular, the existence of the distribution $\CZ$ (that is,
of the vector fields $Z_a$ and $Y^b_a$ fulfilling the above three properties)
must be checked (and guessed) case by case. 
In~\cite{fmpz2,fmt} some GZ
systems, obtained from stationary reductions of the Boussinesq and 
KdV hierarchies, are discussed along these lines.
In the next sections we will apply the scheme herewith outlined
to the Toda lattices.
\end{rem}
\section{The Bihamiltonian approach to Toda lattices}\label{sect:two}
The phase space of the (complex, periodic) Toda lattice 
(see, e.g.,~\cite{Fla})
with $n$ sites (particles) is the manifold
$\CM=(\CC^*)^n\times \CC^n$ parametrized by the Flaschka coordinates
$\{a_i,b_i\}_{i=1,\ldots,n}$. We endow it with the Poisson pencil $P_\la$
defined as follows (see, e.g., \cite{MoPi} and references cited therein). 
It associates with the one--form
$\sum_k (\alpha_k d a_k+\beta_k d b_k)$ the vector field 
$\sum_k ({\dot{a}}_k\partial_{a_k}+{\dot{b}}_k\partial_{b_k})$ 
according to the rule
\begin{equation}\label{eq:pla}
\begin{array}{l}
\dot{a_k}=a_k((b_k-\lambda)\beta_k-(b_{k+1}-\lambda)\beta_{k+1}
+a_{k-1}\alpha_{k-1}-\alpha_{k+1}a_{k+1})\\
\dot{b_k}=(b_k-\lambda)(a_{k-1}\alpha_{k-1}
-a_{k}\alpha_{k})+a_k\beta_{k+1}-a_{k-1}\beta_{k-1}
\end{array}
\end{equation}
where the cyclicity condition
$(\cdot)_{k+n}=(\cdot)_k$
is implicitly assumed.
We write the matrix expression of $P_\la=P_1-\la P_0$
in the $3$--particle case, the $n$--particle case being easily
generalized from this example:
\begin{equation}\label{eq:p13}
P_\la=\left [\begin {array}{cccccc} 0&-a_{{1}}a_{{2}}&a_{{1}}a_{{3}}&a_{{1}}(
b_{{1}}-\la)&-a_{{1}}(b_{{2}}-\la)&0\\\noalign{\medskip}&0&-a_{{2}}
a_{{3}}&0&a_{{2}}(b_{{2}}-\la)&-a_{{2}}(b_{{3}}-\la)\\
\noalign{\medskip}&&0&-a_{{3}}(b_1-\la)&0&a_{{3}}(b_{{3}}-\la)
\\\noalign{\medskip}&&&0&a_{{1}}&-a_{{3}
}\\\noalign{\medskip} &&*&&0&a_{{2
}}\\\noalign{\medskip}&&&&&0\end {array}\right ]
\end{equation}
According to the GZ scheme, we study the kernel of $P_\la$.
We have to solve the equations 
\begin{equation*}\begin{array}{l}
(b_k-\lambda)\beta_k-(b_{k+1}-\lambda)\beta_{k+1}
+a_{k-1}\alpha_{k-1}-a_{k+1}\alpha_{k+1}=0\\
(b_k-\lambda)(a_{k-1}\alpha_{k-1}-a_{k}\alpha_{k})
+a_k\beta_{k+1}-a_{k-1}\beta_{k-1}=0
\end{array}\end{equation*}
With algebraic manipulations (see~\cite{Meu98}),
it can be traded for the system of equations
\begin{equation}\label{eq:syscas}
\begin{array}{l}
(b_k-\lambda)\beta_k+a_{k-1}\alpha_{k-1}+a_k\alpha_k=L_1\\
(a_k\alpha_k)^2+a_k\beta_k\beta_{k+1}-L_1\alpha_k=L_2
\end{array}\end{equation}
where $L_i$ are $\ZZ_n$--invariant functions. 
Setting $L_1=1, L_2=0$, and introducing the variables 
\[
h_k=\frac{\beta_{k+1}}{\alpha_k},
\]
we obtain the following Riccati type equation:
\begin{equation}\label{eq:chareq}
h_k h_{k+1}=(b_{k+1}-\lambda)h_k+a_k.
\end{equation}
\begin{prop}
The {\em characteristic equation}~\rref{eq:chareq} admits a solution
$h_k$ which is a Laurent
series in the parameter $\la$ of the form $h_k=\la+\sum_{j=1}^\infty
h_{k,j}\lambda^{-j}$. The Laurent coefficients $h_{k,j}$
can be computed by recurrence as functions of the variables $\{a_i,b_i\}$.
The product 
$C=h_1\cdots h_n$ of the components of any 
solution of~\rref{eq:chareq} is a Casimir
function of the Poisson pencil $P_\lambda$. 
\end{prop} 
Notice that, once the characteristic equation is solved, 
the one--forms in the
kernel of $P_\la$ can be easily
computed (by recurrence) solving the system
\[
\left\{ \begin{array}{l}
h_k\alpha_k+a_k\beta_k=1\\
                        \alpha_k h_k=\beta_{k+1}\end{array}\right.
\qquad k=1,\ldots, n,
\]
which is equivalent to the system~\rref{eq:syscas} with  $L_1=1,L_2=0$.

This method allows us to find Casimirs of $P_\la$  that are
{\em Laurent series} in $\la$. According to the GZ scheme~\cite{GZ93}, 
however, we should better look for 
{\em polynomial} Casimirs of $P_\la$. They can
be found linearizing the Riccati equation~\rref{eq:chareq} as follows.

Setting  $h_k=\mu \psi_k/\psi_{k-1}$, we transform 
equation~\rref{eq:chareq} into the linear system
\begin{equation}
\mu^2\psi_{k+1}-\mu(b_k-\lambda)\psi_k-a_k\psi_{k-1}=0,
\end{equation}
where $\mu$ is related to the Casimir $C$ via $C=\mu^n$.
In matrix form we have $\CL\mathbf{\psi}=0$, where
\begin{equation}\label{eq:laxmat}
\CL=\left[\begin{array}{ccccc}
\mu(b_1-\lambda)&-\mu^2&0& &a_n\\
a_1&\mu(b_2-\lambda)&-\mu^2&\ddots&\\
0&a_2&\ddots&\ddots&0\\
&\ddots&\ddots&\mu(b_{n-1}-\lambda)&-\mu^2\\
-\mu^2&0&&a_{n-1}&\mu(b_n-\lambda)\\
\end{array}\right].
\end{equation}
This is how the classical Lax matrix of the Toda lattice can be introduced
into the game in the GZ \bih\  point of view. 
We remark that, since the Riccati equation~\rref{eq:chareq} 
admits solutions,  so does the linear system
$\CL\mathbf{\psi}=0$. 
So, taking into account the cyclicity of $\CL$, we arrive at
\begin{prop}
The {\em spectral curve\/} of the problem, 
$\det(\CL)=0$, is a quadratic polynomial in the Casimir $C$, 
\begin{equation}
  \label{eq:spcur}
\det(\CL)=-C^2+\Haa{1}(\lambda)C+\Haa{2} .
\end{equation}
Thus, both $\Haa{1}(\lambda)$ and $\Haa{2}$ are {\em polynomial} 
Casimirs of $P_\lambda$. 
In particular,
\[
\Haa{2}=(-1)^{n+1}a_1\cdots a_n
\]
is a {\em common} 
Casimir of $P_1$ and $P_0$, and $\Haa{1}(\lambda)$ has the form
\[
\Haa{1}(\lambda)=(-1)^{n}\lambda^n+\sum_{j=1}^n \Haa{1}_j\la^{n-j}.
\]
\end{prop}
It can be easily realized that
$\Haa{1}_j=(-1)^j \sigma_j^n(b_1,\dots,b_n)$+lower order terms in the  $b_j$, 
where $\sigma_j^n$ is the $j$--th elementary symmetric polynomial 
in $n$ letters.
So, the Hamiltonian functions
$(\Haa{2},\{\Haa{1}_j\}_{j=1,\ldots,n})$ are functionally
independent and the previous 
proposition provides another proof of the fact~\cite{GZ99} that the
periodic Toda lattice with $n$ particles 
is a complete GZ manifold of type $\{1,2n-1\}$.

\medskip
We end this section with a remark which 
frames the {\em open\/} Toda lattice within this scheme.
It is well known that the open Toda lattice can be obtained form the periodic
one by pulling one particle to infinity, that is, in the Flaschka coordinates,
by letting one of the $a$ coordinates, say $a_n$, attain the $0$ value.
The phase space of the
(complex) open Toda lattice with $n$ particles is thus the 
manifold $\wid{\CM}=(\CC^*)^{n-1}\times \CC^n$ parametrized by 
reduced Flaschka coordinates
$\{a_1\ldots, a_{n-1},b_1,\ldots,b_n\}$.
The Poisson pencil $\wid{P_\la}$  of the open case
can be obtained from the periodic one  by means of the following trick. 
Let $\widetilde{\CM}\supset \CM$ be the manifold 
obtained from the phase space of the periodic lattice adjoining the ``boundary
component'' defined by $a_n=0$.
The Poisson pencil defined by~\rref{eq:pla}, being polynomial in the 
Flaschka coordinates,  
extends naturally to a Poisson pencil $\widetilde{P}_\la$ on 
the extended manifold $\widetilde{\CM}$.
The phase space $\wid{\CM}$ of the open case can be identified
with the zero set of the common Casimir $\Haa{2}$  of $\widetilde{P}_1$ and
$\widetilde{P_0}$, which, obviously enough, is still given by 
$\Haa{2}=(-1)^{n+1}a_1\cdots a_n$.
Then
$\widetilde{P}_\la$ can be restricted to $\wid{\CM}$, and its 
restriction is the Poisson pencil of the open Toda lattice.
In practice, its matrix representation in the reduced coordinates is obtained
by the matrix representation of the periodic Poisson pencil
\rref{eq:p13} 
deleting the $n$--th row and column, and
setting $a_n=0$ in the resulting
matrix. 
The Lax matrix of the open Toda Lattice (as well as the Hamiltonian functions) 
is obtained simply by setting $a_n=0$ in the Lax matrix~\rref{eq:laxmat} of the
periodic problem. In particular, the single polynomial 
Casimir of the Poisson pencil of the open lattice is obtained as
${\wid H}^{(1)}=\Haa{1}{}_{\vert_{a_n=0}}.$ 
The open Toda lattice is thus a complete GZ manifold of type $2n-1$. 

\section{Separation of variables}\label{sect:three} 
In this Section we will show that the Toda lattice fits the scheme
described in Subsection~\ref{sect:2_1}. We will follow the path of
Section~\ref{sect:two}, considering at first the periodic lattice, and then
stating the suitable changes to be done in the
open case.

The periodic Toda lattice is a GZ
manifold of dimension $2n$ and
type $\{1,2n-1\}$. Thus the rank
of the transversal distribution
$\CZ$ must be $2$, and the dimension of the
reduced PN manifold $2n-2$. We divide the procedure 
outlined in Subsection~\ref{sect:2_1} in three 
steps.\medskip\\
{\bf Step 1}.
The transversal
vector
fields $Z_1$ and $Z_2$. 
\begin{prop}
The vector fields $Z_1=\del_{b_n}$ and $Z_2=\del_{a_n}/(a_1\cdots
  a_{n-1})$ satisfy
\begin{equation}
\mbox{Lie}_{Z_1}P_\lambda=Z_1\wedge
Y_{1,1}\
,\qquad 
\mbox{Lie}_{Z_2}P_\lambda=Z_1\wedge
Y_{2,1}\
,
\end{equation}
with $Y_{1,1}=a_{n-1}\del_{a_{n-1}}-a_{n}\del_{a_{n}}$ and 
$ Y_{2,1}=-\del_{b_1}/(a_1\cdots  a_{n-1})$.
\end{prop}
This property is proven making use of the
standard formulas for the Lie derivative of a
bivector.\medskip\\
{\bf Step 2}. The action of $Z_i$ on the Casimirs and the $\lambda_j$ 
coordinates.\medskip\\
To discuss this issue it is useful to recall the expression of the second
Casimir
$\Haa{2}=(-1)^{n+1}
a_1\cdots
a_n$
and
to
expand $\mbox{det}(\CL)$ with respect to the 
last column to get:
\begin{equation}\label{eq:detexpansion}
\mbox{det}(\CL)=\mu(b_n-\lambda)\wid{L}_{n,n}+\mu^2\wid{L}_{n,n-1}
+(-1)^{n+1}a_n \wid{L}_{n,1},
\end{equation}
where $\wid{L}_{i,j}$ are the determinants of the suitable minors 
of $\CL$. 
Taking into account the specific form of these minors one 
can easily see that it holds
\begin{prop} The second Lie derivatives of $\Haa{1}$ and $\Haa{2}$
with respect to $Z_i$ vanish. Furthermore, 
$\mbox{Lie}_{Z_1}(\Haa{2})=0$ and $
\mbox{Lie}_{Z_2}(\Haa{2})=1$, so that the matrix 
$G_a^b=\mbox{Lie}_{Z_a}(\Haa{b}(\lambda))$
 introduced in Subsection~\ref{sect:2_1}
has the form
\begin{equation}
G(\la)=\left(\begin{array}{cc} \mbox{Lie}_{Z_1}(\Haa{1})&0\\
\mbox{Lie}_{Z_2}(\Haa{1})&1\end{array}\right).
\end{equation}
Thus the $\lambda_j$ coordinates are the roots of the monic degree
$(n-1)$ polynomial $ \mbox{Lie}_{Z_1}(\Haa{1})
=\del_{b_n} \det(\CL)=\wid{L}_{n,n}$.
\end{prop}
{\bf Step 3}. The action of the \vefi\
$Y$ and the $\mu_j$ coordinates.\medskip\\
We have to consider the vector field
$Y=Y_{1,1}=a_{n-1}\del_{a_{n-1}}-a_{n}\del_{a_{n}}$,
and to discuss its
action on the exact eigenvectors of $N^*$.
According to the discussion following Proposition \ref{prop:g}, to
construct such an eigenvector we must find a vector $(1,\rho(\la))$
such that $(1,\rho(\la))G(\la)=0$ for $\la=\la_j$. Since this vector
is simply given by $(1,0)$, we have that $\Haa{1}$ is an exact 
eigenvector of $N^*$, and this is true, for all $r$, for 
$Y^r(\Haa{1})$ as well. In order to build a {\em normalized\/} exact
eigenvector, we have to analyze a bit
further the terms in the expansion~\rref{eq:detexpansion} of the 
determinant of $\CL$. Actually, one 
has that:
\begin{enumerate}\item $\wid{L}_{n,n}$ is {\em independent} of $a_n$ and
  $a_{n-1}$;
\item $a_n\wid{L}_{n,1}=\Haa{2}+C K_1$ where $K_1$ is {\em linear} 
in $a_n$ and does not depend on $a_{n-1}$ and $\mu$;
\item $\mu^2\wid{L}_{n,n-1}=C^2+C K_2$ where $K_2$ is {\em linear} 
in $a_{n-1}$ and does not depend on $a_{n}$ and $\mu$.
\end{enumerate}
Thanks  to the linearity properties of $K_j$,
we have that
${Y}(\Haa{1})={Y}(K_1+K_2)$
satisfies the recursion property
$Y^3(\Haa{1})=Y(\Haa{1})$. This
ensures that the function 
\[
F=\log(Y(\Haa{1})+Y^2(\Haa{1}))
\]
satisfies $Y(F)$=1, and, according to
Theorem \ref{prop:y},
is the desired generator of the $\mu_j$ coordinates. 
We notice that, due to the cyclic nature of the periodic 
Toda system, the pair $Z_1,Z_2$ of deformation vector fields is by no means  
unique; other admissible  pairs can be obtained via 
a cyclic permutation of the  indices (with a corresponding change in the
vector field $Y$).
It would be interesting to compare the DN coordinates considered here
with the separation variables for Toda systems used in, e.g.,
\cite{FMcL,Ku97,Sktoda}.

Finally, we state the corresponding results for the open Toda lattice.
We have to look for a single
deformation vector field, say $Z$; it is still given by 
$\del_{b_n}$; the vector field
$Y$ now is given by $a_{n-1}\del_{a_{n-1}}$; the
recursion relation on $Y({\wid H}^{(1)})$ 
closes at the first step, $Y^2({\wid H}^{(1)})=Y({\wid H}^{(1)})$, 
and the generating function can be taken as
$\wid F=\log\left(Y({\wid H}^{(1)})\right)$.

\section{Example: the three--particle case}\label{sect:four}

We close the paper with some explicit expressions
for the three--particle case. The Poisson pencil is written in
equation~\rref{eq:p13}. The Lax matrix is given by
\begin{equation} \CL=\left(  \begin {array}{ccc} \mu\,\left
        (b_{{1}}-\lambda\right )
&-{\mu}^{2}&a_{{3}}\\\noalign{\medskip}a_{{1}}&\mu\,\left (b_{{2}}-\lambda
\right )&-{\mu}^{2}\\\noalign{\medskip}-{\mu}^{2}&a_{{2}}&\mu\,\left (
b_{{3}}-\lambda\right )\end {array}\right).
\end{equation}
The spectral curve is
\[
-C^2-(\la^3+H_0\la^2+H_1\la+H_2)C+K=0,
\]
with
\begin{eqnarray*}
H_0&=&-(b_{{1}}+b_{{3}}+b_{{2}}),\qquad
  H_1=a_{{2}}+b_{{1}}b_{{3}}+a_{{1}}+b_{{2}}b_{{3}}+a_{{3}}+b_{{1}}b_{{2}},\\
H_2&=&-(b_{{1}}b_{{2}}b_{{3}}+a_{{1}}b_{{3}}+b_{{1}}a_{{2}}+a_{{3}}b_{{2}}),
\qquad K=a_{{1}}a_{{2}}a_{{3}},\quad C=\mu^3.\\\end{eqnarray*}
There are two nontrivial flows, given by:
\begin{equation}
X_1=\left\{\begin{array}{l} \dot a_1=a_{{1}}b_{{2}}-a_{{1}}b_{{1}}
\\ \dot b_1=a_{{3}}-a_{{1}}
\\\mbox{and cyclic permutations}\end{array}\right.
\quad X_2=\left\{\begin{array}{l} \dot a_1=a_{{1}}b_{{1}}b_{{3}}+a_{{1}}
a_{{3}}-a_{{1}}b_{{2}}b_{{3}}-a_{{1}}a_{{2}}
\\ \dot b_1=a_{{1}}b_{{3}}-a_{{3}}b_{{2}}
\\\mbox{and cyclic permutations}\end{array}\right.
\end{equation}
The transversal \vefi s are $Z_1=\del_{b_3}$ and
$Z_2=\del_{a_3}/a_1 a_2$, and we have $Y=a_2\del_{a_2}-a_3\del_{a_3}$.
The DN coordinates can be found as follows. The roots of the
polynomial
\[
{\mbox{Lie}}_{Z_1}(\Haa{1}(\la))=\del_{b_3}(H_0\la^2+H_1\la+H_2)
=-\la^2+(b_1+b_2)\la-(b_1 b_2+a_1)
\]
are $\la_1$ and $\la_2$. Then $\mu_1$ and $\mu_2$ are given by the
function
\[
F=\log\left(Y(\Haa{1})+Y^2(\Haa{1})\right)=\log(2a_2\la-2a_2b_1)
\]
evaluated at $\la=\la_1,\la_2$.

For the open case, the Poisson pencil can be computed from~\rref{eq:p13}, 
according to the procedure outlined at the end of Section~\ref{sect:two}:
\begin{equation}
P_\la^{open}=
\left [\begin {array}{ccccc} 0&-a_{{1}}a_{{2}}&\left (b_{{1}}-\lambda
\right )a_{{1}}&\left (\lambda-b_{{2}}\right )a_{{1}}&0
\\\noalign{\medskip}a_{{1}}a_{{2}}&0&0&\left (b_{{2}}-\lambda\right )a
_{{2}}&\left (\lambda-b_{{3}}\right )a_{{2}}\\\noalign{\medskip}\left 
(\lambda-b_{{1}}\right )a_{{1}}&0&0&a_{{1}}&0\\\noalign{\medskip}
\left (b_{{2}}-\lambda\right )a_{{1}}&\left (\lambda-b_{{2}}\right )a_
{{2}}&-a_{{1}}&0&a_{{2}}\\\noalign{\medskip}0&\left (b_{{3}}-\lambda
\right )a_{{2}}&0&-a_{{2}}&0\end {array}\right ],
\end{equation}
and the spectral curve is the rational curve
\[ \begin{split}
\mu^3=C&=-{\lambda}^{3}+\left (b_{{2}}+b_{{1}}+b_{{3}}\right )
{\lambda}^{2}-\left (b_{{1}}b_{{2}}+a_{{1}}+b_{{2}}b_{{3}}+b_{{1}}b_{{
3}}+a_{{2}}\right )\lambda\\&
\qquad\qquad +b_{{1}}b_{{2}}b_{{3}}+a_{{1}}b_{{3}}+b_{{1}
}a_{{2}}.\end{split}
\]
The separation variables can be constructed, {\em mutatis mutandis}, as in
the periodic case.
\subsection*{Acknowledgments} G.F. and M.P. would like to thank the
organizers of NEEDS99 for the opportunity to present these results there,
and for financial support to their participation. 
This work was partially supported by the M.U.R.S.T.
and the G.N.F.M. of the Italian C.N.R.


\begin{thebibliography}{10}

\footnotesize

\bibitem{Ha1} M. R. Adams, J. Harnad, J. Hurtubise, {\em Darboux
Coordinates and Liouville-Arnold Integration  in Loop Algebras.} 
Commun. Math. Phys. {\bf 155} (1993), 385--413.

\bibitem{fmp2} G. Falqui, F. Magri, M. Pedroni, in preparation.

\bibitem{fmpz2}
G. Falqui, F. Magri, M. Pedroni, J. P. Zubelli, 
{\em  A Bi-Hamiltonian theory for  stationary KdV flows
and their separability},
in preparation.

\bibitem{fmt} G. Falqui, F. Magri, G. Tondo, {\em Bihamiltonian systems and
    separation  of variables: an example from the Boussinesq hierarchy.} 
solv-int/9906009, to appear in {\sl Theor. Math. Phys}.

\bibitem{Fla} H. Flaschka, {\em Integrable systems and torus actions\/}.
In: Lectures on Integrable Systems (O. Babelon et al., eds.), World Scientific,
1994, pp.\ 43-101.

\bibitem{FMcL} H. Flaschka, D. W. McLaughlin, {\em Canonically
Conjugate Variables for the Korteweg--de Vries Equation and the 
Toda Lattice with Periodic Boundary Conditions.} Progress Theor. Phys.
{\bf 55} (1976), 438--456.

\bibitem{GZ93} I. M. Gel'fand, I. Zakharevich,
{\em On the local geometry of a bi-Hamiltonian structure.}
In: The Gel'fand Mathematical Seminars 1990-1992
(L. Corwin et al., eds.), Birkh\"auser, Boston, 1993, pp. 51--112.

\bibitem{GZ99}
I. M. Gel'fand, I. Zakharevich,
{\em Webs, Lenard schemes, and the local geometry of \bih\ Toda and Lax
  structures.} math-ag/9903080, to appear in {\sl Selecta Mathematica}.

\bibitem{KM} Y. Kosmann--Schwarzbach, F. Magri, {\em Poisson--Nijenhuis
    structures.} Ann. Inst. Poincar\'e (Phys. Theor.) {\bf 53} 
(1990), 35--81.

\bibitem{Ku97}
V. B. Kuznetsov,
{\em Separation of variables for the ${D}_n$-type periodic
{T}oda lattice.} {J. Phys. A} {\bf 30} (1997), {2127--2138}.

\bibitem{Ma90} F. Magri, {\em Geometry and soliton equations}. In: {\sl La 
M\'ecanique Analytique de Lagrange et son h\'eritage}, Atti
Acc. Sci. Torino Suppl. {\bf 124} (1990), 181--209.

\bibitem{MR} J.E. Marsden, T. Ratiu, {\em Reduction of Poisson Manifolds.}
\lmp{11}{1986}{161--169}.

\bibitem{Meu98} A. Meucci, {\em The \bih\ route to the Discrete Sato
    Grassmannian}. Ph. D. Thesis, Dipartimento di Matematica dell'Universit\`a
  di Milano, 1998.

\bibitem{MoPi}
C. Morosi, L. Pizzocchero,
{\em R-Matrix Theory, Formal {C}asimirs and the Periodic {T}oda Lattice.}
J. Math. Phys. {\bf 37} (1996), 4484--4513.

\bibitem{Sktoda}
E. K. Sklyanin, {\em The quantum {T}oda chain}.
In: {Nonlinear equations in classical and quantum field theory
             (Meudon/Paris, 1983/1984)},
Lecture Notes in Phys. 226, 
Springer, Berlin, 1985, pp.\ 196--233.

\bibitem{Sk} E. K. Sklyanin, {\em Separations of variables:
new trends.} Progr. Theor. Phys. Suppl. {\bf 118} (1995), 35--60.
 
\end{thebibliography}
\end{document}